\documentclass{article}
\usepackage{spconf,amsmath,graphicx,verbatim}
\usepackage{multirow}
\usepackage{amsfonts,amsmath,bm}

\title{Improved Multi-Stage Training of Online Attention-based Encoder-Decoder Models}
           
\name{Abhinav Garg, Dhananjaya Gowda, Ankur Kumar, Kwangyoun Kim, Mehul Kumar, Chanwoo Kim}
\address{Speech Processing Lab, AI Center, Samsung Research, Korea}

\begin{document}
%
\maketitle
\begin{abstract}
 In this paper, we propose a refined multi-stage multi-task training strategy to improve the performance of online attention-based encoder-decoder (AED) models.
 A three-stage training based on three levels of architectural granularity namely, character encoder, byte pair encoding (BPE) based encoder, and attention decoder, is proposed.
 Also, multi-task learning based on two-levels of linguistic granularity namely, character and BPE, is used.
 We explore different pre-training strategies for the encoders including transfer learning from a bidirectional encoder.
 Our encoder-decoder models with online attention show $\sim$\textbf{35\%} and $\sim$\textbf{10\%} relative improvement over their baselines for smaller and bigger models, respectively.
 Our models achieve a word error rate (WER) of \textbf{5.04\%} and \textbf{4.48\%} on the Librispeech test-clean data for the smaller and bigger models respectively after fusion with long short-term memory (LSTM) based external language model (LM).
\end{abstract}
\begin{keywords}
Attention based encoder-decoder models, online attention, multi-stage training, multi-task learning
\end{keywords}
\section{Introduction}
\label{sec:intro}
Recently, attention-based encoder-decoder (AED) models have gained popularity for developing end-to-end neural network based automatic speech recognition (ASR) systems~\cite{google_sota,returnn18,park2019specaugment}.
One of the primary advantages of AED models is that the language information is tightly coupled into the decoder, obviating the need for an external language model (LM).
AED models have been shown to perform better than other end-to-end models, namely, connectionist temporal classification (CTC) and recurrent neural network transducer (RNN-T) models~\cite{prabhavalkar2017comparison}.
With large amounts of transcribed data ($>$1000 hrs), the AED models perform even better than the conventional DNN-HMM systems  ~\cite{google_sota, park2019specaugment}.
This makes the AED models one of the best candidates for both server-side and on-device ASR applications.
However, most of the best performing AED models use a bidirectional long short-term memory (BLSTM) based encoder and attend to the entire input sequence, which makes them not suitable for online decoding.

Conventional HMM-DNN based ASR systems which use unidirectional long short-term memory (ULSTM), time-delay neural network (TDNN) or convolutional neural network (CNN) based architectures, as well as other end-to-end architectures namely, CTC and RNN-T models,  are capable of online decoding.
However, both the HMM-DNN and CTC models need an external LM to achieve good performance, especially for large vocabulary ASR systems.
RNN-T models also need a separate predictor network to capture the language information. RNN-T models with small footprints which are also capable of online decoding have been recently shown to perform extremely well for on-device applications~\cite{google_rnnt_streaming}.
AED models with monotonic attention have also been proposed earlier for online decoding~\cite{raffel2017online, mocha}.
A hard monotonic attention based encoder-decoder was first proposed in~\cite{raffel2017online}.
A much improved soft monotonic chunkwise attention (MoChA) based model was proposed in~\cite{mocha}.
However, the performance of these monotonic attention models lags behind their full attention~\cite{attn2} counterparts, as well as other model architectures capable of online decoding.
In this paper, our primary focus is to improve the performance of encoder-decoder models with online attention.

Online end-to-end models (CTC or attention-based) with deep neural architectures, when trained with random initialization, often have had difficulties in converging well\cite{TS_training_3stage, yu2018multistage}.
Several initialization strategies have been proposed to avoid this problem and improve model convergence.
In~\cite{othere2einit}, it was shown that using a pre-training strategy improves the generalization capability and hence performance of models in several seq2seq problems such as machine translation and abstractive summarization. 
Initializing a word-based CTC model with a pre-trained phone-based CTC model was found to be useful in~\cite{init1}.
Similarly, multi-task learning on hierarchical models has also been found to be effective, as in~\cite{subword1}. Using teacher-student transfer learning, to train a CTC based online ULSTM student from a BLSTM teacher model has been explored in~\cite{TS_training_3stage}.

Motivated by these pre-training and initialization techniques, we propose a multi-stage training strategy for online AED models.
In terms of contribution: (a) We propose a new multi-stage training strategy for improving the accuracy of online AED models. 
(b) We explore three different methods for training the ULSTM based encoder models at two levels of linguistic granularity, character and byte pair encoding (BPE)~\cite{sennrich2015neural_bpe} units.
(c) For each of the above three methods, we propose different pre-training strategies that help the models converge better. These strategies are not limited to our methods, but can also be applied to any LSTM based ASR model in general.

This work is organized in the following manner. Section 2 gives an overview of attention based encoder-decoder models. Section 3 describes our strategy. Section 4 contains experimental details and results. And we conclude in Section 5.
\vskip -0.15in
 \begin{figure}[t]
 \centering
  \includegraphics[]{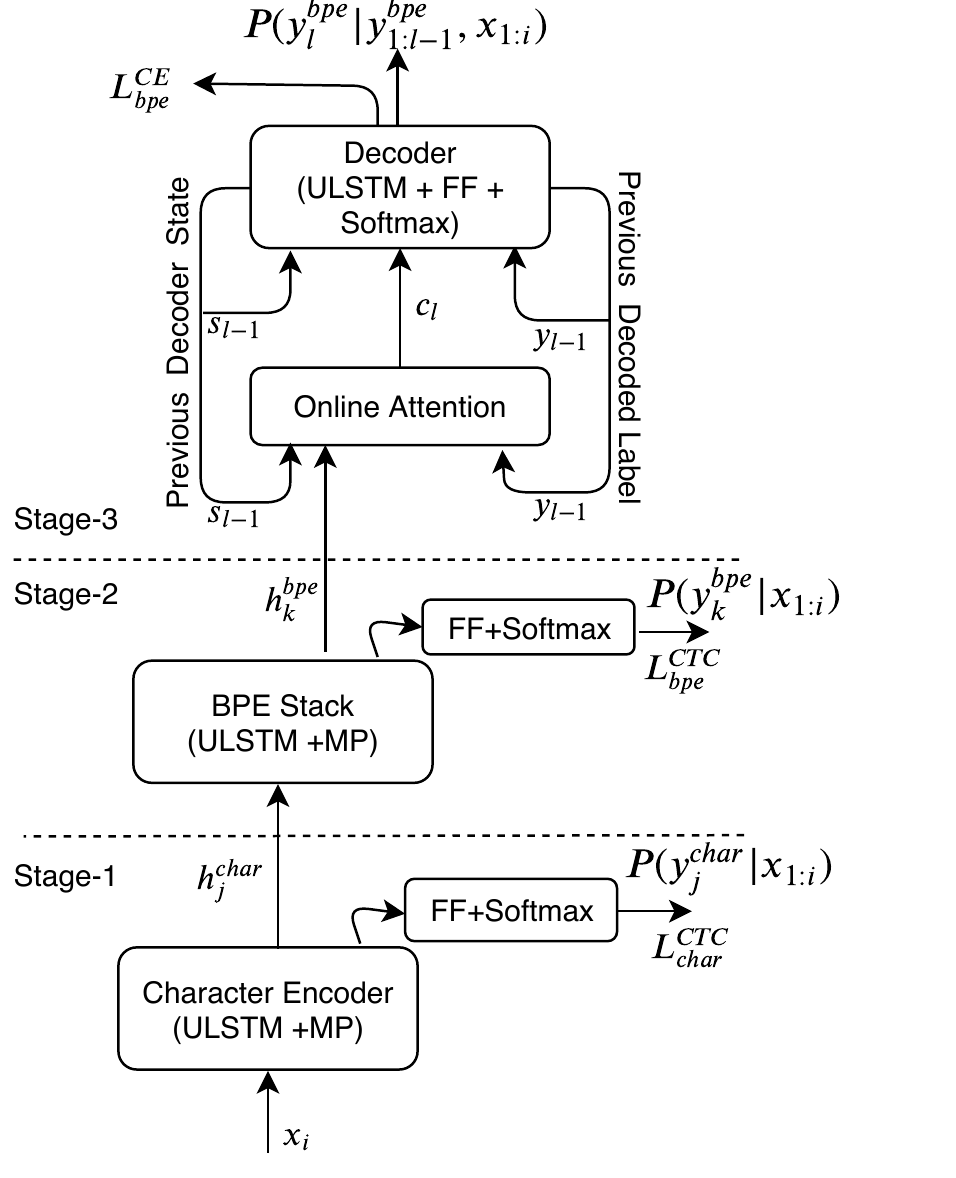}
  \caption{Block schematic of the proposed multi-stage multi-task character-to-BPE online AED model.}
  \label{CBoA}
\end{figure}
\section{Attention based encoder-decoder models}
\label{sec:aed}
A block schematic of the multi-stage multi-task character-to-BPE (C2B) online AED model proposed in this paper is shown in Fig.~\ref{CBoA}.
The model consists of three distinct stages, a character encoder, a BPE stack, and an attention decoder.
The character encoder consists of several ULSTM layers with an optional maxpool (MP) layer in-between. 
It converts an input sequence $\bm{x}=\{x_i\}_{i=1}^{T}$ of length $T$ into a sequence of hidden character embeddings $\bm{h}^{char}=\{h_j^{char}\}_{j=1}^{T_c}$ of length $T_c$, and $j=\lfloor{i/T_s}\rfloor$ where $T_s$ denotes the overall time subsampling (TS) factor introduced by the MP layers.
The BPE stack consists of a few additional ULSTM layers again with an optional MP layer in-between.
This converts the input character embeddings into a sequence of BPE embeddings $\bm{h}^{bpe}=\{h_k^{bpe}\}_{k=1}^{T_b}$ of length $T_b$.
A single feed-forward (FF) projection layer followed by a softmax is used on top of the character encoder and BPE stack to convert the embeddings ($\bm{h}^{char}$ and $\bm{h}^{bpe}$) into probability distributions ($P(y_j^{char}|x_{1:i})$ and $P(y_k^{bpe}|x_{1:i})$) over the corresponding character and BPE target labels, respectively, as shown in the figure.
The character encoder and the BPE encoder are trained using two different CTC losses on character and BPE targets ($L^{char}_{CTC}$ and $L^{bpe}_{CTC}$), respectively~\cite{Graves:2006:CTC:1143844.1143891}.
The CTC loss function for any chosen target labels $lab$ (can be either $char$ or $bpe$ in our case) is given by 
\begin{align}
    L^{lab}_{CTC}=-log \sum_{\bm{\pi}\in A(\bm{y}^{lab*},\bm{h}^{lab})}{\prod_{j} P(\pi^{lab}_j|x_{1:i})}.
\end{align}
Here, $\bm{\pi}=A(\bm{y}^{lab*},\bm{h}^{lab})$ denotes one of all possible alignments between the ground-truth target sequence $\bm{y}^{lab*}$ and the encoder embeddings $\bm{h}^{lab}$ either by repeating the output labels or by inserting a blank symbol.

An MP layer has no trainable parameters, but helps primarily in reducing the length of an input sequence by a predetermined pooling factor.
It has been observed that reducing the input sequence length progressively through the encoder helps in better convergence and accuracy of the AED models~\cite{returnn18, attn2, povey2016purely}.
In this work, we use an overall time reduction factor of 2 for character encoder, and an additional reduction factor of 4 in the BPE stack, taking the overall reduction factor for the BPE encoder to 8.
An MP layer wherever used has a reduction factor of 2 along the time-dimension unless specified otherwise.
In this work, we use BPE based sub-word units as the final target labels for the AED model.
Character labels are used as intermediate targets in a multi-task training strategy.

The attention decoder stage consists of an attention block and a decoder block.
The attention block computes a context vector $c_l$ based on the encoder embeddings $\{h_{1:k}^{bpe}\}$, previous predicted label $y_{l-1}$, and the previous decoder state $s_{l-1}$. Where, $l$ is the target label produced at output after processing inputs $x_{1:i}$.
In this paper, we use MoChA for the attention block to ensure online or streaming decoding~\cite{mocha}.
The decoder consists of one or more ULSTM layers followed by a few fully connected FF layers and a softmax layer.
The decoder uses the attention context vector, the previous state of the decoder, and the previous decoded label to generate the BPE label probabilities $P(y_l^{bpe}|y^{bpe}_{1:l-1}, x_{1:i})$ at the softmax output.
The AED model is trained with a cross-entropy (CE) loss $L_{bpe}^{CE}$ between these label probabilities and the ground truth target labels given by
\begin{align}
    L_{bpe}^{CE}=-\sum_{l=1}^{L}{\log( P(y_l^{bpe*}|y^{bpe*}_{1:l-1}x_{1:i}))},
\end{align}
where $*$ denotes ground truth BPE target labels and $L$ is the length of the target sequence.

\vskip -0.15in
\section{Multi-stage multi-task training of online attention models}
\label{sec:ms_mr_c2b}
Training CTC based encoder models have often been found to be difficult with random initialization~\cite{subword1}.
Initializing these models with weights pre-trained based on frame-wise cross-entropy using prior alignment between the input and output sequences has shown better convergence of these models~\cite{tied_triphone,alignment_from_CE}.
In~\cite{TS_training_3stage}, a teacher-student transfer learning has been used to improve the convergence as well as the performance of character based CTC encoder models.
In the case of AED models, it has been shown that a carefully designed layerwise pre-training strategy helps the models to converge better~\cite{returnn18}.
Similarly, a joint multi-task training using a CTC loss on the encoder and a CE loss on the attention decoder was proposed to improve the convergence and accuracies of AED models~\cite{joint_ctc_ce,kim2017joint}.

In view of this, we propose a multi-stage multi-task training strategy to train our character-to-BPE (C2B) online AED models.
The three different stages of our proposed training strategy are outlined below.
In our work, \textit{pre-initialization} would mean borrowing weights from an already well-trained model and \textit{pre-training} would mean modifying some aspects of the network architecture during the initial part of the training. 
\vskip -0.10in
\subsection{Stage-1: Training the CTC-Char encoder}
\label{sec:CTC-Char}
In the first stage, we train the CTC based character encoder (CTC-Char) model.
We use a deep stack of ULSTM layers to predict character sequence $\bm{y}_{char}$ given the input feature sequence $\bm{x}$. We use one maxpool layer after the first ULSTM layer to reduce the encoder output length by a factor of 2. 
The character encoder output or embeddings for any given input $\bm{x}$ is given by
\begin{align}
    {\bm h}_{char} = {\cal L}^{u}_{n}\circ\cdots\circ {\cal L}^{u}_{2}
    \circ {\cal P} \circ {\cal L}^{u}_{1}(\bm{x}) \label{eq1}
\end{align}
where ${\cal L}^{u}_{i}$ denotes the $i^{th}$ ULSTM layer, and ${{\cal P}}$ denotes a maxpool layer with a pooling factor of 2. 
The character encoder embeddings $\bm{h}_{char}$ are passed through a linear projection or FF layer followed by a softmax layer which captures the distribution of the character labels along with the blank symbol used in the computation of CTC loss.
This network is trained to minimize the CTC loss between the softmax output and the ground truth label sequence $y_{char}$. 

In order to improve the convergence of this model, we apply a layer-wise pre-training strategy moving from a higher time reduction factor to a lower factor, similar to ~\cite{returnn_e2e}. 
We start with three ULSTM layers and we then incrementally add a new ULSTM layer approximately every $1/3^{rd}$ of an epoch. 
Also, we use fixed MP layer after the first ULSTM layer and a floating MP layer before the last ULSTM layer, in order to maintain an overall time-reduction factor of 4 during the pre-training.
The floating MP layer is removed after adding all the ULSTM layers reducing the overall time-reduction factor to two.
Starting the training with all the ULTSM layers at once with a reduction factor of 2 leads to sub-optimal results, around 10\% higher relative error as compared to our approach. 
Using higher reduction factor (8 or 16) is not feasible for character based CTC encoders as the input sequences tend to become shorter than the target sequences.
\vskip -0.10in
\subsection{Stage-2: Training the CTC-BPE encoder}
\begin{figure*}[tbh]
 \centering
 \vskip -0.48in
  \includegraphics[width=15cm]{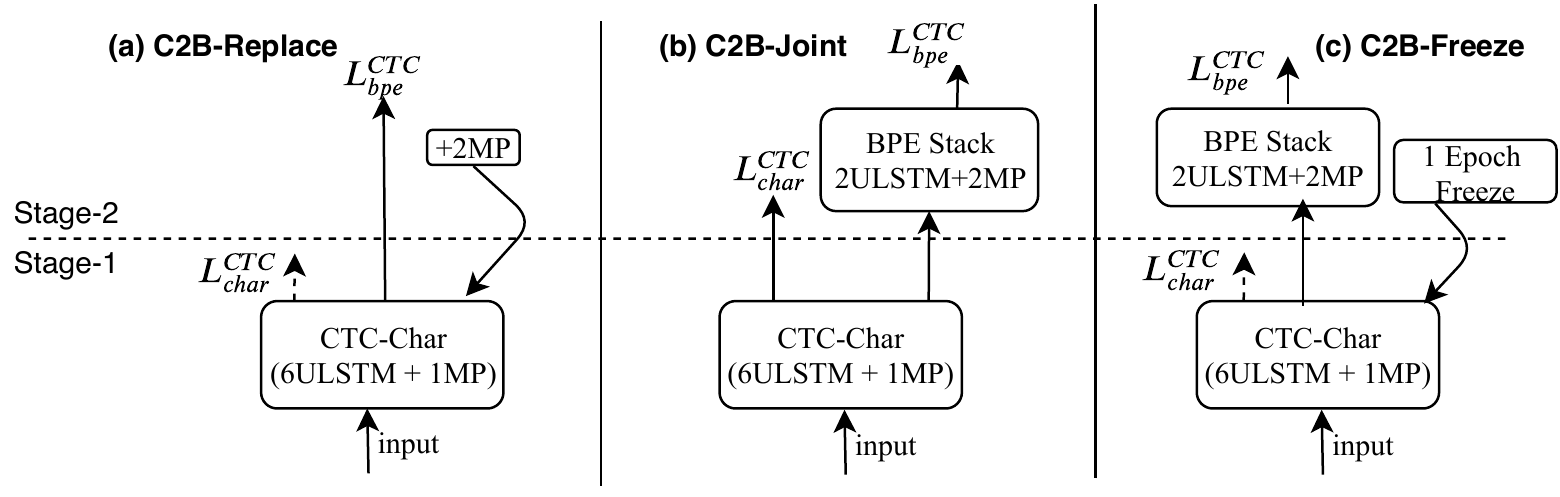}
  \vskip -0.18in
  \caption{Overview of methods to convert character encoder to BPE encoder. The dashed arrows for $L_{char}^{CTC}$ in C2B-Replace and C2B-Freeze indicate that they will be removed during Stage-2.}
  \vskip -0.18in
  \label{Compare_Method}
\end{figure*}

In Stage-2, we start with the character encoder model trained in Stage-1, and convert it into a BPE encoder (CTC-BPE) model by using one of the three methods depicted in Fig.~\ref{Compare_Method}.
\vskip -0.10in

\subsubsection{C2B-Replace: Char-to-BPE encoder by replacing losses}
In this method, we convert the pre-trained character encoder to BPE encoder by replacing the character CTC loss with BPE CTC loss.
Two additional MP layers are added after the $2^{nd}$ and $3^{rd}$ ULSTM layers, taking the overall time-reduction factor to 8 including a factor of 2 from the character encoder.
The BPE encoder output is given by
\begin{align}
    {\bm h}_{bpe} = {\cal L}^u_{n} \circ \cdots\circ {\cal P} \circ {\cal L}^u_{3} \circ {\cal P} \circ {\cal L}^u_{2} \circ {\cal P} \circ {\cal L}^{u}_{1}(\bm{x}). \label{eq2}
\end{align}
Similar to the pre-training strategy of the character encoder, we start with a high time-reduction factor of 32 using additional MP layers and gradually removing them one by one to a final factor of 8.
\subsubsection{C2B-Joint: Char-to-BPE encoder with joint losses}
In this method, we convert the pre-trained character encoder to a BPE encoder by adding a BPE stack containing two ULSTM layers on top of character model and two additional MP layers to increase the overall reduction factor to 8.
The BPE encoder output for this model is given by
\begin{align}
    {\bm h}_{bpe} = {\cal L}^u_{bpe2} \circ {\cal P} \circ {\cal L}^{u}_{bpe1} \circ {\cal P} \circ (\bm{h}_{char}).
\end{align}
The model is trained with the joint losses of character CTC and BPE CTC as shown in Fig.~\ref{Compare_Method}(b).
There was no additional pre-training strategy applied for this method.

\subsubsection{C2B-Freeze: C2B-Joint with an initial freeze on character encoder}
In this method, the conversion from character to BPE encoder is exactly similar to the C2B-Joint method, except for the absence of character CTC loss and freezing the character encoder for approximately one epoch after adding the additional BPE stack.
An additional pretraining strategy was used by increasing the pooling factors of last two MP layers in the BPE stack from 2 to 4, making the overall reduction factor as 32 at the start of training.
The overall reduction factor is gradually reduced to 8, by reducing the factors of the last two MP layers from 4 to 2, one at a time after half and one epoch of training, respectively.
The BPE encoder output for this model is given by
\begin{align}
    {\bm h}_{bpe} = {\cal L}^u_{bpe2} \circ {\cal P} \circ {\cal L}^{u}_{bpe1} \circ {\cal P} \circ {\cal F}(\bm{h}_{char}),
\end{align}
\vskip -0.10in
where ${\cal F}$ denotes a temporary freeze operation on the pre-trained character model.


\subsection{Stage-3: Training the attention-decoder}
In this stage, we use the BPE encoder trained from Stage-2 and build an end-to-end attention system by attaching a decoder with MoChA attention on top of it. Our decoder has a single ULSTM layer. We also use attention feedback and a maxout layer similar to to~\cite{returnn18}. This whole network is now trained with the cross-entropy loss between decoder output and the ground truth BPE target labels. 
\section{Experiments and results}
\vskip -0.08in
\subsection{Datasets and training}
All our experiments are reported on the publicly available 1000 hours Librispeech corpus~\cite{LibriSpeech}. 
The {\em dev} and {\em test} sets consist of {\em clean} and {\em other} subsets. For our ULSTM encoder experiments (CTC-Char and CTC-BPE), we use combined dev set consisting of {\em dev-clean} and {\em dev-other} to evaluate our models.
Our character vocabulary consists of 29 labels, all upper-case English alphabets, an unknown symbol, an end of sentence (eos) symbol, and space symbol to denote word boundaries. For BPE, we use a vocabulary of size 10K for all our experiments.
The word error rate (WER) results are reported for our online AED models on the {\em test-clean} and {\em test-other} subsets.

We use 40-dimensional MFCC features extracted with librosa~\cite{librosa}, which are then combined into variable batch sized chunks so as to keep the number of frames processed per training step constant. 
We use 6 ULSTM layers for the CTC-Char models each with 256 units, and a dropout of 0.3~\cite{dropout}. We also use batch normalization after each ULSTM layer of the encoder~\cite{batch_norm}. 
Our decoder consists of a single ULSTM layer with 1024 units.
We use RETURNN framework for all our experiments~\cite{returnn16,returnn18,abadi2016tensorflow}.
In all our experiments we use Adam optimizer~\cite{Adam} with an initial learning rate of 0.0008 and learning rate scheduling using a cross-validation set.  
We also use linear learning rate warm-up~\cite{google_sota} along with gradient clipping.
For generating the best hypothesis during the test time, we use beam search with beam size 12. 

\begin{table}[t]
\caption{Performance (CER in \%) on combined {\em dev} set of our CTC-Char model trained using C2B-Joint multi-stage method, as against the baseline CTC-Char models trained with and without pre-training (PT) as in Section~\ref{sec:CTC-Char}, and BLSTM-to-ULSTM (B2U) teacher-student (TS) approach.}
\vskip -0.05in
\label{CTC-Char}
\begin{center}
\begin{small}
\begin{sc}
\begin{tabular}{|l|c|}
\hline
Model & CER(\%) w/o LM \\
\hline
CTC-Char Baseline  & 18.38 \\
CTC-Char Baseline + PT & 16.40 \\
B2U TS training  & 16.22 \\
CTC-Char from C2B-Joint   & \textbf{13.61} \\
\hline
\end{tabular}
\end{sc}
\end{small}
\end{center}
\vskip -0.30in
\end{table}
\begin{table*}[t]
\caption{Performance of the CTC-BPE models trained with different architectures and strategies in terms of BER(\%) on the combined {\em dev} set. DNC denotes `Did not converge'.}
\vskip -0.10in
\label{CTC-bpe}
\begin{center}
\begin{small}
\begin{sc}
\begin{tabular}{|l|c|c|c|c|c|}
\hline
\multirow{2}{*}{Model} &  \multicolumn{2}{c|}{Loss} & \multirow{2}{*}{Init} & \multirow{2}{*}{Pre-training} & \multirow{2}{*}{BER(\%) w/o LM}\\\cline{2-3}
      & $L_{CTC}^{char}$ & $L_{CTC}^{bpe}$ & & & \\ \hline
\multicolumn{6}{|l|}{Single-stage trained models}  \\\hline
\multirow{3}{*}{Baseline (8L ULSTM + MP)} 
                                & no  & yes & Random & no  & DNC \\
                                & yes & yes & Random & no & 33.15 \\
                                & no  & yes & Random & PT  & 26.96 \\\hline
\multicolumn{6}{|l|}{Multi-stage trained models}  \\\hline
\multirow{2}{*}{C2B-Replace}    & no  & yes & CTC-Char & no & 28.61\\
                                & no  & yes & CTC-Char & PT & 24.91\\\hline
\multirow{2}{*}{C2B-Joint}      & yes & yes & CTC-Char & no & 24.08 \\
                                & yes & yes & CTC-Char & PT & DNC \\\hline
\multirow{3}{*}{C2B-Freeze}     & no  & yes & CTC-Char & Freeze & 26.76 \\
                                & no  & yes & CTC-Char & PT & DNC \\
                                & no  & yes & CTC-Char & Freeze+PT &{\bf 23.63}\\\hline
\end{tabular}
\end{sc}
\end{small}
\end{center}
\vskip -0.25in
\end{table*}
\vskip -0.95in
\subsection{Effect of multi-stage training on CTC-Char model}
In this section, we study the effect of our multi-stage training on the CTC character encoder model.
For this, we use the CTC-Char model obtained after Stage-2 training of the C2B-Joint strategy, after removing the BPE stack, as our candidate model.
The CTC-Char models from the other two strategies (C2B-Replace and C2B-Freeze) could not be evaluated owing to the removal of the CTC-Char losses during Stage-2 training.
We compare our model against two baselines. 
(1) A 6 layer ULSTM baseline trained without any pre-training. 
(2) A 6 layer ULSTM baseline trained using similar pre-training strategy as ours in Section~\ref{sec:CTC-Char}.

We also compare our multi-stage training strategy with previously proposed teacher-student transfer learning based approach~\cite{TS_training_3stage}. 
We train a BLSTM based CTC character encoder, exactly similar to training a ULSTM CTC character encoder as outlined in Section~\ref{sec:CTC-Char}), as the teacher model.
The teacher model had 256 BLSTM units in each layer in each direction, and achieved a character error rate (CER) of 7.6\%. We then use 
frame-wise Kullback-Liebler (KL) divergence based teacher-student transfer learning to train the student model.

Table~\ref{CTC-Char} shows the performance of different character encoder models in terms of CER on the combined {\em dev} set.
Just with pre-training, the vanilla or baseline CTC-Char model achieves 10\% relative improvement over the baseline. 
Our proposed multi-stage approach performs $\sim$26\% better than the baseline without PT, and $\sim$17\% better than the baseline with PT.
During the training of C2B-Joint, character loss increases initially but ultimately settles down to a much lower value (26\% relative) due to guidance from the BPE loss.
Also, our model performs $\sim$16\% better than model trained using teacher-student training without the need of any teacher model.
\subsection{Performance of CTC-BPE encoder model}
The performance of various CTC-BPE encoder models in terms of BPE label error rate (BER) on the combined {\em dev} set is given in Table~\ref{CTC-bpe}.
As a baseline, we use an encoder model with 8 layer ULSTM stack and 3 MP layers, trained in a single stage. 
When trained with a single BPE CTC loss, random initialization and without any pre-training, the model failed to converge.
This may be due to the lack of sufficient gradient flow in the backward direction for the update of weights in an otherwise deep network. 
When the same model is trained with an additional character CTC loss after $6^{th}$ layer, it converged to a BER of $\sim$33.15\%. 
We obtain the best results of $\sim$27\% BER when we trained this baseline model with only the BPE CTC loss, but with a pre-training similar to that outlined in Section~\ref{sec:ms_mr_c2b}. We use this model as the baseline for our CTC-BPE experiments.\\
\phantom{x}\hspace{0.4cm} To show the effectiveness of pre-training techniques we also calculate BER for models trained with and without pre-training (only applicable for C2B-Replace and C2B-Freeze).
We tried pre-training C2B-Joint with a strategy similar to C2B-Freeze but it did not converge for us. A more carefully crafted pre-training might be required for well convergence of joint loss models having CTC-Char initialization. Also, not freezing the CTC-Char weights leads to non-convergence as the initial epochs of CTC-BPE training adversely affect the well trained CTC-Char weights. Table~\ref{CTC-bpe} shows the corresponding results. For each method, pre-training gives better performance as compared to normal training. While training baseline with BPE CTC loss without pre-training does not converge, training similar architecture, C2B-Freeze with pre-initialization converges well even without pre-training suggesting that CTC-Char weights are indeed a very good initializer for CTC-BPE models. We achieve $\sim$10.6\% relative improvement using C2B-Joint, $\sim$7.6\% relative improvement using C2B-Replace and $\sim$12.35\% relative improvement using C2B-Freeze over the baseline.

We note that each of our methods has some advantage over the others. C2B-Joint has one big advantage that it also improves the performance of CTC-Char online model. Joint loss between character and BPE helps both of them to converge to a better optimum. For applications where we need character outputs, we can simply discard the two additional layers and just use the character stack.
C2B-Replace has an obvious advantage of being smaller in size and performing comparably to others. C2B-Freeze is the best performing method amongst the three.

\subsection{Performance of online C2B-AED models}
\vskip -0.08in
\subsubsection{Baseline}
\vskip -0.08in
For comparison with our models, we train equivalent models without pre-initialization (i.e. random initialization). These models contain 8 ULSTM layers as the encoder, with a time reduction of two after each of the first three ULSTM layers. We use MoChA online attention and decoder architectures the same as mentioned before. Along with that, we use layer-wise pre-training similar to those in Section~\ref{sec:ms_mr_c2b}, where we start with an initial reduction factor of 32 and decrease it incrementally to 8 after one epoch.

For better convergence we try different training techniques including (1) Training with only CE loss (2) Training with combined CTC and CE loss~\cite{joint_ctc_ce}  (3) Using combined loss for the first pass over the data (i.e. pre-training stage) and training with CE loss afterward.
CTC loss is computed between the outputs of BPE encoder and target labels. And CE loss is computed between the outputs of decoder and target labels.
We compare the performance of these baselines in Table~\ref{baseline}. As we can see using CTC during initial epochs helps as it ensures better initial convergence for ULSTM encoder weights.
\vskip -0.15in
\begin{table}[t]
\caption{WER(\%) for our baseline AED online models trained with different strategies. Here PT-CTC refers to "CTC loss used only during pre-training"}
\label{baseline}
\begin{center}
\begin{small}
\begin{sc}
\begin{tabular}{|l|c|c|}
\hline
Model & \multicolumn{2}{c|}{WER(\%) w/o LM} \\\hline
          & clean & other \\\hline
CE Baseline & 11.56 &28.17\\
CE + CTC Baseline&10.20  & 26.88 \\
PT-CTC + CE Baseline  & \textbf{8.35} & {\bf 22.87}\\
\hline
\end{tabular}
\end{sc}
\end{small}
\end{center}
\vskip -0.28in
\end{table}
\subsubsection{Our proposed C2B online AED model}
\label{our-exp}
\vskip -0.08in
Table~\ref{Mocha WER} shows the performance of our C2B online AED model as compared to the baselines mentioned above.
We use the C2B-Joint model from Stage-2 with a BER of 24.08\% to initialize our Stage-3 AED model primarily for its simplicity in training and good performance. 
Also, following our observations during training of baseline models, we discard the CTC losses used in C2B-Joint BPE encoder.

We also combine our models with an external RNN language model (RNN-LM) obtained from~\cite{returnn_e2e}. This RNN-LM was trained on officially available 800M word dataset of LibriSpeech and has a perplexity score of $\sim$65.9. We use shallow fusion~\cite{shallow_fusion} for combining RNN-LM and our model. For test-clean, we achieve $\sim$30\% relative improvement without external LM and over 35\% relative with external LM. Our final C2B online AED model has only a total of around 20M parameters. And achieves \textbf{5.04} WER on LibriSpeech. This is the best WERs reported in literature on Librispeech corpus for online attention based encoder-decoder models, to the best of our knowledge.
\vskip -0.25in
\subsubsection{Experiments with bigger model}
\vskip -0.08in
We also try our approaches on bigger models with ULSTM cell size as 1024 instead of 256 having around 4 times more parameters ($\sim$ 86M parameters). The baseline for 1024 models has same architecture and training as the baseline for 256 MoChA model used in table~\ref{Mocha WER} except for the cell size.

We also explored initializing lower layers of the encoder with a model trained on an auxiliary loss. Specifically, we trained a 3 layer ULSTM model with frame level CE loss and phone target for each frame. These layer weights were then used to initialize lower layers of 5 layer ULSTM model which had an MP layer (pooling factor of 2) after the third ULSTM layer. The model is trained with CTC loss and phone sequence as the target. We now use these weights for initializing encoder which is 7 layer ULSTM model with MP layers after the third ULSTM layer (pooling factor of 2) and fifth ULSTM layer (pooling factor of 4). We also attach decoder and MoChA attention. The model is then trained with CE loss with the sequence of BPE as the target. We also use pre-training where we add a MP layer (pooling factor of 4) after the sixth ULSTM layer till half the data is seen.  Frame level target and overall phone sequence for given audio were obtained from HMM/DMM systems using Montreal Forced Aligner~\cite{mcauliffe2017montreal} and Kaldi toolkit~\cite{kaldi}. We call these experiments as FCE experiments.

Results for above 1024 sized models are summarized in table~\ref{1024}. Here also our strategy achieves over 7\% relative improvement over the baseline without using LM. And about 8.4\% relative improvement after doing shallow fusion using the same external LM  as Section~\ref{our-exp}.

\vskip -0.20in
\begin{table}[t]
\caption{WER(\%) comparison of proposed C2B AED online model with baseline}
\vskip -0.10in
\label{Mocha WER}
\begin{center}
\begin{small}
\begin{sc}
\begin{tabular}{|l|c|c|c|c|}
\hline
\multirow{2}{*}{Model} & \multicolumn{2}{c|}{WER(w/o LM)} & \multicolumn{2}{c|}{WER (w/ LM)}\\\cline{2-5}
 & clean & other & clean & other \\\hline
CE+CTC Baseline & 10.20  & 26.88 & 7.88  & 24.82 \\
C2B AED         & {\bf 7.24}   & {\bf 20.80} & {\bf 5.04} & {\bf 16.61} \\\hline

\end{tabular}
\end{sc}
\end{small}
\end{center}
\vskip -0.3in
\end{table}
\begin{table}[t]
\caption{WER(\%) comparisions of bigger AED models (having encoder cell size 1024) with other architectures}
\vskip -0.18in
\label{1024}
\begin{center}
\begin{small}
\begin{sc}
\begin{tabular}{|l|c|c|c|c|}
\hline
\multirow{2}{*}{Model} & \multicolumn{2}{c|}{WER(w/o LM)} & \multicolumn{2}{c|}{WER (w/ LM)}\\\cline{2-5}
 & clean & other & clean & other \\\hline
CE+CTC Baseline  & 6.78 & 20.12 & 4.89 & 16.23\\
C2B AED          & {\bf 6.25} & {\bf 18.33} & {\bf 4.48} & {\bf 15.93}\\\hline
FCE (w/ ext align) & {\bf 6.19} & {\bf 17.76} & {\bf 4.35} & {\bf 15.17} \\ \hline
\multicolumn{5}{|c|}{Other competitive online architectures} \\\hline
TDNN-LSTM \cite{capio} & - & - &3.65&8.79\\\hline
CTC/ASG \cite{jasper} & 3.86 & 11.95 &2.95&8.79\\
\hline
\end{tabular}
\end{sc}
\end{small}
\end{center}
\vskip -0.40in
\end{table}
\section{Conclusion}
\vskip -0.10in
We present an effective multi-stage training strategy for online attention encoder-decoder models that achieves significant improvements over it's baseline. All three stages of our 3-staged training strategy are themselves online end-to-end system at three different levels of granularity. Our strategy achieves significant improvements for all the stages involved. 
The performance of our online C2B AED models are the best WERs reported on Librispeech corpus for online attention based encoder-decoder models, to the best of our knowledge.

\bibliographystyle{IEEEbib}
\bibliography{strings,refs}

\end{document}